\documentclass{article}
\usepackage{times}
\usepackage{amsfonts}
\usepackage{amsthm}
\usepackage{amsmath}

\usepackage[all]{xy}

\newbox\ncintdbox \newbox\ncinttbox 
\setbox0=\hbox{$-$} \setbox2=\hbox{$\displaystyle\int$}
\setbox\ncintdbox=\hbox{\rlap{\hbox
    to \wd2{\hskip-.125em\box2\relax\hfil}}\box0\kern.1em}
\setbox0=\hbox{$\vcenter{\hrule width 4pt}$}
\setbox2=\hbox{$\textstyle\int$}
\setbox\ncinttbox=\hbox{\rlap{\hbox
    to \wd2{\hskip-.125em\box2\relax\hfil}}\box0\kern.1em}

\newcommand{\stroke}{\mathbin|}     

\def\proof{\noindent {\bf proof:\ }}
\def\endproof{\vrule height 0.5em depth 0.2em width 0.5em}
\newtheorem{lemma}{Lemma}[section]

\newtheorem{thm}[lemma]{Theorem}
\newtheorem{proposition}[lemma]{Proposition}

\theoremstyle{definition}

\newtheorem{remark}[lemma]{remark}

\newcommand{\C}{\mathbb{C}}       
\newcommand{\R}{\mathbb{R}}       
\newcommand{\Z}{\mathbb{Z}}       

\newcommand{\bea}{\begin{eqnarray}}
\newcommand{\eea}{\end{eqnarray}}
\newcommand{\beas}{\begin{eqnarray*}}
\newcommand{\eeas}{\end{eqnarray*}}
\newcommand{\be}{\begin{equation}}
\newcommand{\ee}{\end{equation}}

\def\<#1,#2>{\langle#1\stroke#2\rangle} 

\def\Z{{\mathbb Z}}
\def\C{{\mathbb C}}
\def\R{{\mathbb R}}

\def\ei{\,\,\,{\rm l} \!\!\!\!\!\; 1\,}

\def\CA{{\cal A}}

\def\CM{{\cal M}}

\def\CZ{{\cal Z}}

\def\CQ{{\cal Q}}

\def\CH{{\cal {H}}}

\def\oh{\frac{1}{2}}

\begin{document}
\begin{center}
\vspace*{10mm} {\bf \Large Generally covariant quantum mechanics
on noncommutative configuration spaces}

\vspace{1cm}

{\large Tom\'{a}\v{s} Kopf\footnote{T.Kopf@math.slu.cz} \\
 Silesian
University, Matematical Institute , Na Rybn\'{\i}\v{c}ku 1, 74601
Opava, Czech Republic\\
\begin{center}
and \end{center}
Mario Paschke\footnote{Mario.Paschke@uni-muenster.de}\\
Institut f\"ur Mathematik, Einsteinstrasse 62 \\ 48149 M\"unster,
Germany}

\end{center}

\vspace{1.5cm} \noindent

\begin{abstract}
We generalize the previously given algebraic version of ``Feynman's
proof of Maxwell's equations'' to  noncommutative configuration
spaces. By doing so, we also obtain an axiomatic formulation of
nonrelativistic quantum mechanics over such spaces, which, in
contrast to most examples discussed in the literature, does not rely
on a distinguished set of coordinates.
\\
We give a detailed account of several examples, e.g., $C^\infty
(\CQ)\otimes M_n(\C)$ which leads to
nonabelian Yang-Mills theories, and of noncommutative tori $T_{\theta}^d$. \\
Moreover we, examine models over the Moyal-deformed plane
$\R^2_\theta$. Assuming the conservation of electrical charges, we
show that in this case the canonical uncertainty relation $[x_k,
\dot{x}_l] = ig_{kl}$  with metric $g_{kl}$ is only consistent if
$g_{kl}$ is constant.
\end{abstract}

 \vspace{0.5cm}

\section{Introduction}

In 1990, F. J. Dyson published a proof \cite{Dyson} of R. P.
Feynman showing that the only interactions compatible with the
canonical commutation relations of quantum mechanics are the
electromagnetic ones, with the electric and magnetic fields
satisfying the homogeneous Maxwell equations. This proof was
generalized to the setting of gauge fields by C. R. Lee \cite{Lee}
and an attempt towards a relativistic theory was made by S.
Tanimura \cite{Tani}. Further considerations appear also in a
review by J.F Carinena, L.A. Ibort, G.Marmo and A.Stern
\cite{Fein-revi}.

The idea of Feynman's proof of Maxwell theory allowed in
\cite{Paschke} a global and completely operational formulation of
scalar quantum mechanics in close correspondence with the axioms
of spectral geometry as given in \cite{C1} in which the
configuration space $Q$ is described by the algebra of smooth
coordinate functions $C_0^{\infty}(Q)$ vanishing at infinity and
represented on the Hilbert space of the quantum particle.

This paper extends these results to noncommutative geometry by
providing in Section \ref{ncs} the needed framework and working out
as a standard example the Moyal-deformed plane
$\mathbb{R}^2_\theta$. We shall then present a number of almost
commutative examples in Section \ref{examples}: The case of
$C_0^{\infty}(Q)\otimes M_n(\mathbb{C})$ in general leading to
Quantum mechanics in interaction with a gauge field (for suitable
$n$ interpretable as a spin connection). Then, in particular and in
more detail the two-point model followed by a new example of a
noncommutative deformation of the torus  $T^2$. The usual
noncommutative torus $T^2_\theta$ is investigated in Section
\ref{nc-tor}.

As will become clear from these examples, our axioms may be used
to define quantum mechanical systems over rather generic
noncommutative configuration spaces, even though it is for most
cases very hard to compute the most general Hamiltonian compatible
with our axioms.

In any case, the noncommutative uncertainty relation turns out to be
very restrictive as it essentially states that the Hamiltonian $H$
is ``a differential operator of second order'' in the sense of
Connes' Noncommutative Geometry. The usage of this property of $H$
as guiding principle might, at first sight, appear rather ad hoc. It
turns out, however, that it should be viewed as a ``nonrelativistic
approximation''. It is, in fact, quite remarkable that the
noncommutative case appears more naturally in the context of
relativistic quantum field theory \cite{PaVE}. The difficulties with
a relativistic one-particle quantum mechanics on noncommutative
spaces will be given in \cite{relQM}.

However, to our point of view the main achievement of this paper is
to clarify which $pre-C^*$-algebras, faithfully represented on a
Hilbert space,  may be good candidates to serve as noncommutative
configuration spaces for quantum mechanics: Those that are small
enough to allow enough room for functionally independent velocities
and at the same time large enough to essentially generate together
with these velocities the full algebra of observables. To illustrate
the difficulties we shall give examples of algebras that are chosen
inappropriately in Section  \ref{ncs}.


\section{Scalar quantum mechanics}

Scalar quantum mechanics as given in \cite{Paschke} is a global
formulation of quantum mechanics on a configuration manifold
$\mathcal{Q}$ which is in an algebraic setting fully captured by its
algebra $\mathcal{A}=C^{\infty}_0(\mathcal{Q})$ of smooth functions
vanishing at infinity. It consists of a family $(\mathcal{A}_t, t\in
\mathbb{R})$ of unitary representations of the algebra $\mathcal{A}$
satisfying the following rather general postulates:

\begin{enumerate}
\item {\bf Localizability:} The representations $\mathcal{A}_t$
contain in their representation space $\mathcal{H}$ a dense finite
projective module $\mathcal{H}_{\infty}$  (It is isomorphic to
representations of $C^{\infty}_0(\mathcal{Q})$ on the space
$\mathcal{H}=L^2(\mathcal{Q},E)$ of square integrable sections of
a complex line bundle $E$ over $\mathcal{Q}$ vanishing at
infinity).

\item {\bf Scalarity:} At every time $t$, the commutant
$\mathcal{A}_t^{'}$ of $\mathcal{A}_t$ (the set of all operators
commuting with $\mathcal{A}_t$) is just $\bar{\mathcal{A}_t}$, the
closure of $\mathcal{A}_t$ in the weak topology:

\begin{align}\label{scalarity}
\mathcal{A}_t^{'}=\bar{\mathcal{A}_t}
\end{align}

\item {\bf Uncertainty relation:} The time evolution, i.e., the
dependence of the representation $a_t$ of any $a\in \mathcal{A}$
is smooth in the parameter $t$ with respect to the strong topology
and satisfies:
\begin{align}\label{2ndOrder}
i[\mathcal{A}_t,\dot{\mathcal{A}_t}]&\subset \mathcal{A}_t &
\text{for all $t\in\mathbb{R}$}
\end{align}
\item {\bf Positivity and nontriviality:} For all $a_t \in
\mathcal{A}_t$
\begin{align}
-i[a_t,\dot{a_t}]&\geq 0,\\
\intertext{and equality}
[a_t,\dot{a_t}]&= 0\\
\intertext{for some $a_t \in \mathcal{A}_t$ implies} \dot{a_t}&= 0
\end{align}
\end{enumerate}

This set-up allows for easy algebraic expressions of geometric
concepts.

Linear connections on $E$ along a vector field $X$ are fully
characterized as operators
$\nabla_X:\mathcal{H}_{\infty}\rightarrow\mathcal{H}_{\infty}$
satisfying
\begin{align}
{\nabla_X}^{\ast}&= \nabla_X,\\
\nabla_X (a_t \psi) &= iX(a_t)\psi + a_t\nabla_X\psi,\\
\nabla_{a_t X+b_t Y} &= a_t \nabla_X + b_t \nabla_Y,
\end{align}
for any vector fields $X$, $Y$ and for any
$a_t,b_t\in\mathcal{A}_t$, $\psi\in \mathcal{H}_\infty$. The
freedom in the choice of a connection $\nabla$ is given by a form
$A$ on $\mathcal{Q}$ that relates it to an arbitrarily chosen
reference connection $\partial$.

\begin{align}
\nabla_X = \partial_X + \langle A, X \rangle.
\end{align}

The following facts are derived in \cite{Paschke}
\begin{enumerate}

\item For each time $t$ and for any $a_t, b_t\in \mathcal{A}$, the
expression
\begin{align}
i g_t(da_t, db_t) &= [a_t, \dot{b_t}]
\end{align}
defines an inverse Riemannian metric $g_t$.

Its symmetry follows from the commutativity of $\mathcal{A}$ and
its consistency with the Leibniz rule follows from the commutator
$[\bullet, \dot{b}]$ satisfying the Leibniz rule and from the
symmetry of $g$.

\item There exists (up to a topological obstruction) a Hamiltonian
operator $H$ that generates the time evolution. Thus,
surprisingly, the uncertainty relation (\ref{2ndOrder}) enforces
the time evolution to be unitary, i.e $\| a_t\|$ is independent of
$t$.

\item By (\ref{2ndOrder}) it is bound to be a differential
operator of order two. Together with (\ref{scalarity}), Newton's
law (the expression of second time derivatives in terms of first
and zeroth ones) can be shown to hold. \\
The Hamiltonian has the form
\begin{align}\label{Ham}
H=\frac 1 2 \Delta(g_t, A_t) + \phi_t
\end{align}
where $\Delta(g_t, A_t)$ is the covariant Laplacian associated
with $g_t$, $A_t$ and $\phi_t\in \bar{\mathcal{A}_t}$.
\end{enumerate}

This shows that the postulates of scalar quantum mechanics,
although general in their appearance, lead to a tightly specified
dynamics: A Hamiltonian of {\em second order in spatial
derivatives} of the particular form (\ref{Ham}) and a Newton's law
expressing {\em second time derivatives} in terms of lower ones.

\section{Noncommutative configuration spaces}\label{ncs}
Being formulated in terms of (commutative) operator-algebras of
observables it is tempting to try to generalize the axioms of
scalar quantum mechanics \cite{Paschke} to
generic noncommutative algebras for at least two reasons:\\
First of all, there might be additional quantum numbers like
isospin, which render
the "configuration space" noncommutative.\\
Second, one might ask wether spacetime itself is a noncommutative
space, like the Moyal-deformed plane
\[ [x,y] = i\theta \]
that has been quite fashionable over the last years. In that case,
one should of course try to find a sound basis to set up quantum
mechanical models on such a space.
\\
To our point of view, this in particular means that the physical
content of such a model should be independent of the coordinates we
have chosen to describe it. In other words, the model should not
depend on the choice of generators of the algebra of functions on
the configuration space. This is, however, not the case for many
examples discussed in the literature. Yet, if spacetime is really
noncommutative, then there should be a dynamical mechanism leading
to this noncommutativity and we believe this mechanism to be related
to the quantized gravitational field. As the (classical) dynamics of
the  gravitational field are governed by the demand of general
covariance, i.e., the independence of the chosen coordinate system,
we do therefore not believe that models which heavily violate
general covariance could be of physical relevance. One might of
course argue that, at very small scales there might appear an
anomaly to diffeomorpism invariance and that violations of the
latter , i.e., the appearance of preferred coordinates, may
therefore be visible at such a scale. But even then, it seems to us,
such effects should not be visible in nonrelativistic one-particle
systems. We do not exclude that in many particle systems such
effects might accumulate and thus be rendered observable at
macroscopic scales. Yet, we do not consider
such systems here, but leave their investigation open as a future project. \\

It is usually assumed in the literature that the noncommutativity
parameter $\theta$ in $[x,y]=i\theta$ is a physical parameter that
can be determined experimentally just as $\hbar$ can be measured.
Yet, to us this seems to be a wrong analogy (to quantum mechanics).
The point is that due to Darboux' theorem there do exist {\em
canonical, i.e. uniquely given } momenta $p_i$ to (almost) any
reasonable choice of coordinates $q_i$, such that
$[q_i,p_j]=i\delta_{ij}\hbar$. Thus $\hbar$ does not depend on the
coordinates we have chosen on the configuration space, but only
makes reference to the symplectic structure, which in turn is
uniquely given by the dynamics. In contrast to that, if the
coordinate $x_1$ is given, then there is no canonical choice of the
remaining coordinates $x_k$. Thus, we can always achieve by an
appropriate choice of coordinates that $\theta=1$, say. In fact, the
algebras generated by $[x,y]=i\theta$ are  isomorphic for all
nonvanishing values of $\theta$ and our models will be equivalent
for isomorphic algebras, and hence $\theta$ will not be observable
in the sense of general relativity.  But there is  a crucial
difference if  $\theta$ is vanishing, as we shall see.

\subsection{Axioms for noncommutative scalar quantum mechanics}

\noindent It is well known, that the space of sections of vector
bundles  can equivalently be described as a finitely generated
projective module over $\CA$. This notion, however, is also
well-defined for noncommutative algebras, so we shall replace the
corresponding axiom,
by requiring \\

\noindent {\bf NC-''Localizability''.} {\em For all $t\in
\mathbb{R}$ the representations are finitely generated projective
modules over $\CA$}.

As the methods used in \cite{Paschke} to infer the existence of a
Hamiltonian are not available in
the noncommutative case, we shall for simplicity also require the \\

\noindent {\bf Unitarity of Time-evolution} {\em , i.e., we shall
require the existence of a (possibly time-dependent) selfadjoint Hamiltonian $H$
generating the time evolution $\CA_t$.}\\

If the algebra $\CA$ is noncommutative,  it is in general no longer
possible to require that the commutator of "functions", i.e., of
algebra elements $a_t\in \CA_t $ and
"velocities" $\dot{b}_t\in \dot{\CA}_t$ is an algebra element: \\
Suppose $a_t,b_t \in \CA_t$ and
\[  [a_t,\dot{b}_t]\in \CA_t .\]
Then one has
\[ [a_t, \dot{(b_tc_t)}] = [a_t,b_t]\dot{c}_t +
b_t[a_t,\dot{c}_t]+[a_t,\dot{b}_t]c_t
   + \dot{b}_t[a_t,c_t]
\]
which will not be in the algebra for all $c_t\in\CA_t$, unless
$\dot{b}_t$ is either an algebra element itself, or the factor
$[a_t,c_t]$ vanishes. It is thus still possible to demand that the
commutant of elements of the center , that we shall denote $\CZ_t
:=Z(\CA_t)$, with time
derivatives of algebra elements are in the algebra:\\

\noindent {\bf weak NC-Uncertainty} {\em We assume that
         \[ [z_t, \dot{b}_t] \in \CA_t\]
for all $z_t\in \CZ_t$ and all $b_t\in \CA_t$.} \\

\noindent
\begin{proposition} \label{ncu-lemma}
Suppose the NC-Uncertainty holds. Then \bea
\, [\CA_t, \dot{\CZ}_t] \subset \CA_t \\
\, [\CZ_t , \dot{\CZ}_t] \subset \CZ_t \eea
\end{proposition}

\proof \\
Assume $z_t\in \CZ_t$ and $a_t\in\CA_t$. Then:
\[ [z_t,a_t] = 0 \qquad \Rightarrow \qquad [z_t,\dot{a}_t] = [a_t,\dot{z}_t] \]
(by differentiation with respect to $t$). This proves the statement
$ [\CA_t, \dot{\CZ}_t]
\subset \CA_t$.\\
Consider now additional  elements $w_t\in \CZ_t$. The statement
$[\CZ_t , \dot{\CZ}_t] \subset \CZ_t$ then follows from the Jacobi
identity: \beas 0 = [a_t,[z_t,\dot{w}_t]] +
\underbrace{[z_t,\underbrace{[\dot{w}_t,a_t]}_{\in\CA_t}]}_{=0} +
    [\dot{w}_t,\underbrace{[a_t,z_t]}_{=0}] \qquad\qquad\qquad\endproof
\eeas As we shall see in the next section, the NC-Uncertainty is
in some cases very restrictive. However, in the generic case, the
center of $\CA$ might be too small, or even trivial, so that we
shall need additional assumptions for such cases.

However, we have not yet generalized the axiom of ``scalarity''. A
noncommutative algebra can, of course, not equal its commutant.
However, we might require it to be antiisomorphic to it. Note that
this follows immediately if there exists a cyclic and separating
vector for the algebra $\CA_t$, as is the case for the vacuum
representation of the algebra of local observables of a quantum
field theory. In that case the operator $J$ is related to the
PCT-operator. Thus, in a sense the above requirement would mean that
we should not be able to distinguish a relativistic scalar particle
and its PCT transformation on such a noncommutative spacetime. As
has been pointed out in \cite{PaVE}, this has, in fact to be
expected. We therefore consider our requirement as naturally
motivated from QFT to which a nonrelativistic QM should be an
approximation.

\noindent {\bf NC-Scalarity} {\em We shall assume, that there
exists  for each $t$ an antiunitary operator $J_t$, such that
\begin{equation}
\overline{J_t \CA_t J_t^{-1}} = \CA_t' .  \label{scal}
\end{equation}
} \\
In other words, we shall assume that the algebras $\CA_t \pi_t(\CA)$ are   anti-isomorphic to a dense subse of their commutant. Equivalently
that then means that besides the family of representations $\pi_t$
of the algebra $\CA$ of ``functions'' over the noncommutative
configuration space there is also a family $\pi^o_t= J_t \pi_t^*
J_t^{-1}$ of ``opposite'' representations given, i.e., one has:
\[ \pi^o_t(ab) = \pi^o_t(b) \pi^o_t(a) \qquad\qquad \forall t\in \R ,\quad \forall a_t,b_t\in \CA_t .\]
For brevity, we shall use the notation $\pi_t^o(a) = a^o_t$, thus we
have
\[ [a_t, b_t^o]=0 \qquad\qquad \forall t, \quad \forall a,b\in \CA .\]
In the commutative case one can  choose $J_t$ to be given for all
times $t$ by the complex conjugation of functions, i.e.
$J_ta_tJ_t^{-1} = \overline{a_t}$. In view of this example it
therefore appears natural to assume the {\bf reality-condition}:
\[ \frac{{\rm d}}{{\rm d} t} \left( J_t a_t J_t^{-1} \right)
= J_t \left( \frac{{\rm d}}{{\rm d} t} a_t \right) J_t^{-1} \qquad
\qquad \forall a_t\in \CA_t   \] which is obviously fulfilled in
the classical case. Note that this condition is already slightly
weaker then requiring $J$ to commute with $H$. However, this
condition still can not always be met in the noncommutative case,
as we shall see in the examples. In any case, the property
NC-scalarity now enables us to require a stronger version of the
uncertainty relation:\\

\noindent {\bf strong NC-Uncertainty}: {\em If the above property
of NC-Scalarity  holds, then it is assumed that
\[ [\dot{a}_t, b^o_t] \, \in \, \CA^o_t\otimes\CA_t  \qquad\qquad \forall t, \quad  \forall a_t,b_t \in \CA_t . \]
}

The notation $\CA^o_t\otimes\CA_t $ should be understood as the
set of all operators that can be written in the form $\sum_i
a_{t,(i)} b_{t,(i)}^o $.

Hence, the commutators with time derivatives $\dot{a}_t$ are
required to be derivations on the algebras $\CA^o_t$. Note that
this condition then also implies that $J_t \dot{a}_t J_t^{-1}$ is
a derivation on the algebra $\CA$. If the above reality property
holds, then one also has $\dot{(a^o_t)}=J_t \dot{a}_t J_t^{-1}$
and thus the strong NC-Uncertainty is completely symmetric in
$\CA_t$ and $\CA_t^o$.

\begin{proposition}
Strong NC-Uncertainty implies  weak NC-Uncertainty, i.e
\[ [z_t, \dot{a}_t] \in \CA_t \qquad \qquad \forall z_t \in \CZ_t, \quad a_t\in \CA_t .  \]
\end{proposition}

\proof Since the commutant of $\CA^o_t =  J_t \CA_t J_t^{-1}$
equals $\CA_t$ it is sufficient to prove that $[z_t, \dot{a}_t]$
commutes for all $ z_t\in\CZ_t$ and  $a_t\in \CA_t$ with all
$b^o_t \in \CA^o_t$. Considering the Jacobi-Identity again:
\[ 0 = [[z_t, \dot{a}_t],b_t^o] + [\underbrace{[b_t^o, z_t]}_{=0},\dot{a}_t] +
\underbrace{[\underbrace{[\dot{a}_t,b_t^o]}_{\in
\CA_t\otimes\CA_t^o},z_t]}_{=0} .  \qquad \qquad\qquad \endproof
\]

\begin{proposition}
If $\CA\otimes\CA^{o}=\CZ(\CA)^{'}$ then weak NC-Uncertainty
implies strong NC-uncertainty.
\end{proposition}

\proof \\
To show strong NC-Uncertainty, it is by the assumptions sufficient
to show that $[\dot{a_t},b_t^o]$ is in the commutant of
$\CZ_t(\CA_t)$, i.e.,
\begin{align}
[[\dot{a_t},b_t^o],z_t]&=0& \text{ for all $a_t, b_t\in\CA_t$,
$z_t\in \CZ_t(\CA_t)$}
\end{align}
But this holds due to the Jacobi identity
$$[[\dot{a_t},b_t^op],z_t]=[[\dot{a_t},z_t],b_t^op]$$ and since
$[\dot{a_t},z_t]$ is in $\CA_t$ by weak NC-Uncertainty and thus
commutes with $b_t^o$.
\endproof\\

However, for infinite dimensional algebras $\CA\otimes\CA^o $ will in general only be a (dense) subset of the commutant
of $\CZ$. Yet, for finite dimensional algebras,  weak and  strong uncertainy are equivalent:\\
Let $\CA = \oplus_{k=1}^{K}{M_{n_k}(\mathbb{C})}$ be any finite
dimensional $C^{\ast}$-algebra. Define $P_k = \mathbf{1}_{n_k}$,
i.e.,
\begin{align}
P_k P_i &= \delta_{ik} P_i, &\\
P_k a &\in M_{n_k}(\mathbb{C})& \text{ for all $a\in\CA$,
$k=1,2,\ldots , K$}
\end{align}
The centre of $\CA$ is then given as $$\CZ(\CA) = \{ \sum_k c_k P_k
\mid c_k\in\mathbb{C}\}.$$ It is well known (see, e.g.,
\cite{Paschke-Sitarz}) that the only representation $\pi$ of $\CA$
such that there exists an antiunitary automorphism
$J:\mathcal{H}\rightarrow \mathcal{H}$ with $J\CA J^{-1}=\CA^{'}$ is
given by taking $\mathcal{H}=\CA$ (as a vector space) and
representing $a$ by left multiplication on matrices:
\begin{align}
\pi(a) \psi &= a\psi
\end{align}
$J$ is (up to unitary equivalence) given as
\begin{align}
J \psi &= \psi^{\ast},
\end{align}
where $\psi$ is considered as a block-diagonal matrix. It follows
that $J^{-1}=J$

Then
\begin{align}
a^{op} \psi &=J a^{\ast} J\psi=\psi a
\end{align}
Thus $\CA^{op}$ is left multiplication with $\CA$ Note now that,
with $\mathcal{H}_k = P_k\mathcal{H}$one has that
\begin{align}
\pi(\CA) &: \mathcal{H}_k \rightarrow \mathcal{H}_k &&\text{ for all $k$}\\
\pi^{op}(\CA)=J\pi(\CA)J &: \mathcal{H}_k \rightarrow \mathcal{H}_k
&&\text{ for all $k$},
\end{align}
as they both commute with $P_k$

One then easily proves that
\begin{align}
P_k(\pi(\CA)\otimes\pi^{op}(\CA))P_k \cong
\text{End}(\mathcal{H}_k)=P_k \text{End}(\mathcal{H}) P_k.
\end{align}
Since $\sum_{k}{P_k}=\mathbf{1}_{\mathcal{H}}$, we have

\begin{align}
\sum_{k}{P_k(\pi(\CA)\otimes\pi^{op}(\CA))P_k} &= \pi(\CA)\otimes\pi^{op}(\CA),\\
\intertext{and thus}
\pi(\CZ(\CA)^{'}) &= \{ M:\mathcal{H}\rightarrow\mathcal{H}\mid M P_k=P_k M\} =\\
&= \{ M:\mathcal{H}_k\rightarrow\mathcal{H}_k\} =\\
&= \sum_{k}{P_k(\text{End }\mathcal{H}) P_k} =\\
&= \sum_{k}{P_k(\text{End }\mathcal{H}_k) P_k} =\\
&= \pi(\CA)\otimes\pi^{op}(\CA).
\end{align}

Thus, for finite dimensional algebras, weak NC-Uncertainty and
strong NC-Uncertainty are equivalent. As we shall see in the example of the noncommutative torus, which has a
trivial center, the strong uncertainty is in fact very restrictive for infinite dimensional algebras, however.

\subsection{Heuristic construction for the Moyal-plane}

Let us illustrate the difficulties on the example of the
Moyal-deformed plane $[x_1,x_2] =i\theta$. One might then require
the canonical uncertainty relations \bea [x_k,\dot{x}_l] \in \CA
\label{moy-cure} \eea for the generators $x_k$.  As explained above,
it would then not be true that $[a,\dot{b}]\in\CA$ for generic
algebra elements $a,b\in\CA$. In particular, if one would choose
another set of generators of the algebra, then these new generators
would  not fulfill the canonical uncertainty relation. Thus, models
which are constructed in this way depend on the choice of
generators!
\\
Nevertheless, it is quite instructive to proceed with the
construction of models based on (\ref{moy-cure}).  To this end we
shall use the standard (star-product) representation of the Moyal
deformed  plane on the space $\CA=L^2(\R^2)$ (which is just the
representation of the
algebra on itself). \\
Note that there do exist the  standard derivations
\[ \delta_i (x_j) = \delta_{ij}  \]
on the algebra and hence they are also represented (as hermitean
operators upon multiplying with $i$) on the Hilbert space $\CH$.
Then, as usual one has $[a,\delta_k] = i\delta_k(a) $ for all
$a\in \CA$. The representations of the $\delta_i$ are not in the
algebra. Recall that for any $b\in\CA$ that is not in the center,
$[a,b\delta_k]$ will not be an algebra element (for all $a$). Thus
$b\delta_k$ does not define a derivation on $\CA$.\\

For simplicity we shall only consider derivations of the algebra
of the form
\[ P(a) = \sum_i \zeta_i \delta_i (a) + [A,a] \]
with $\zeta_i\in \mathbb{C}$ and $A\in \mathcal{A}$. (Most
probable all derivations of $\CA$ are actually of this form.
However we are not aware of any proof of this statement.)

\noindent Let us define the non-vanishing operators
$$p_k = \delta_k -\varepsilon_{kl} \,\theta\, x_l $$
Obviously these operators commute with all the algebra elements.
In fact,
one can prove that the commutant $\CA'$ of $\CA$ is generated by $p_1,p_2$.\\

\noindent Let us now return to (\ref{moy-cure}). It then
immediately follows (from the Leibniz rule for commutators), that
for every algebra element (power series in the generators) $a\in
\CA$ one has
\[ [a,\dot{x}_k]  \in \CA . \]
Thus commuting with $\dot{x}_k$ is a derivation on the algebra
$\CA$. If we then allow for $H$ only derivations on $\CA$ of the
form given above, there do thus exist $\zeta_{kl},A_k$ such that
\[ \dot{x}_k = i\sum_l\,\zeta_{kl}\, \delta_l  + A_k  + \omega_k,  \]
where $\omega_k\in \CA'$. We suppress $\omega_k$ in the remainder.
Therefore
\[ [x_l, \dot{x}_k] =  i \zeta_{kl}  +[x_l,A_k] . \]
Note also that, upon differentiating $[x_k,x_l] i\varepsilon_{kl}\theta $ with respect to $t$ we get the
consistency relation:
\[ i\zeta_{kl} + [x_l,A_k] = i\zeta_{lk} + [x_k, A_l].    \]

Recall now the classical case: There we found $\dot{x}_k i\sum_l\, g_{kl}\, \delta_l  + A_k $ , but $g_{kl}$, i.e the
components of the metric being in $\CA$. So here it seems, that
the metric has to be constant. Well, not really, since now the
vector potential $A_k$ renders the uncertainty nontrivial.  As we
shall point out in the section on the noncommutative Torus (where
we obtain a similar result) the term $[x_l,A_k]$ does in fact
correspond to the metric on this space.

In conclusion, one can consistently construct quantum mechanical
models over the Moyal deformed plane. However, these models turn
out to be fairly restricted as compared to the commutative case.

\section{Almost commutative Spaces}\label{examples}
We now turn our attention to simple examples for algebras of the
form $\CA=C^\infty_0(\CQ) \otimes \CA_f $, where $\CA_f$ is a
(semi-simple) finite-dimensional $algebra$. Such algebras do have a
very large center. As we shall see, our NC-Uncertainty condition is
therefore already a very strong (sometimes even too strong)
requirement.

\subsection{$C^\infty_0(\CQ)\otimes M_n(\C)$ and nonabelian gauge theories}
We first consider the algebra $$\CA = C^\infty_0(\CQ)\otimes
M_n(\C).$$ We shall represent it on the Hilbert space $\CH
L^2(\CQ,E)\otimes \C^n $, which corresponds to a "line bundle" over
$\CA$ i.e. it is given (up to closure...) as $\CH = p \CA$ where
$p=p^2\in\CA$ has tr$p=1$. It is well known, that in this case one
has:
\[   \CA' = \overline{\CZ} , \qquad \qquad\qquad \CZ' = \overline{\CA}. \] \\
Choosing generators $T^a$ of $su(n)$ in the fundamental
representation (they then together with the unit matrix do form a
basis in $M_n(\C)$), we thus have (using local coordinates):
\[
[x^k,x^l]  = [x^k, T^a] =  0 \qquad \qquad \qquad [T^a,T^b]  =  i
f^{ab}_c T^c .
\]

If we then demand the NC-Uncertainty
$$ [x^k,\dot{a}]\in \CA \qquad\qquad\qquad \forall a\in \CA, $$
then we immediately get from the above lemma:
\[ [x^k,\dot{x}^l] \in \CZ \qquad\qquad\qquad [\dot{x}^k, T^a] \in \CA. \]
Analogously to the classical case, we can thus infer:
\[ \dot{x}^k = g^{kl} (-i \partial_l - A_l) \]
where $g^{kl}\in \CZ= C^\infty_0(\CQ)$ and $A_l = a_l^*\in \CA$.
Note that $A_l$ thus defines a ${\bf u}(n) = {\bf u}(1) \oplus
{\bf su}(n)$ gauge connection. But then we have
\[ [x^k,\dot{T}^a] = [T^a,\dot{x}^k] = g^{kl}[T^a,A_l] \]
from which we immediately get
\begin{lemma}
There exist hermitean elements $A_0^a\in \overline{\CA} = \CZ' $
such that
\[ \dot{T}^a = -\dot{x}^k [A_k,T^a] + A_0^a  \]
\end{lemma}
Constructing the most general Hamiltonian, that would lead to
these time derivatives, we then obtain:
\begin{thm} \label{generalH}
There exists a metric $g^{kl}$ and a ${\bf u}(n)$-gauge connection
$A_\mu$ ($\mu =0,1,\ldots,d$) on $\CQ$ (respectively on $\CM \CQ\times \R$ such that
\[ H = \oh g^{kl} (-i\partial_k - A_k)(-i\partial_l - A_l)  + A_0 . \]
\end{thm}
Hence we obtain that our system describes a scalar particle that
is (minimally) coupled
to gravity and ${\bf u}(n)$ Yang-Mills Theory.\\
It is quite surprising that our axioms are that restrictive in
this case. In particular, we obtain the nonabelian gauge
invariance just from the NC-uncertainty
relation (that is defined on the center only ! ). \\

\begin{remark}  Let us restrict to $n=2$ in
which case $T^k =\sigma^k$, $k=1,2,3$ and $\CQ=\R^3$. Then one would
also expect to obtain the Pauli-Term
$\frac{e}{m}\vec{B}\vec{\sigma}$ and the fine structure term
$$\frac{e}{4m^2r} \frac{\partial\varphi}{\partial r}
\vec{\sigma}\vec{L} $$ (for radially symmetric scalar potential
$\varphi$)
to appear in $H$. \\
They are, in fact, present: Choose
$A_0=\frac{e}{m}\vec{B}\vec{\sigma}$ to obtain the Pauli term and
$\vec{A}= - \frac {ie} {8 m^2}\vec{\sigma}\times\vec{\nabla}\phi(r)$
 to obtain the fine structure. The physical difference of spinors and $SU(2)-doublets$, i.e. the
correct spin-orbit coupling is, of course, only seen in the relativistic theory.
\end{remark}
\begin{remark}
Since $\CH=L^2(\CQ,E) \otimes \C^n$, we only obtain the ${\bf
su}(n)$ gauge theory on a trivial bundle, while the ${\bf u}(1) $
part might live on an arbitrary line bundle. If one would like to
get gauge theories on arbitrary bundles of rank $n$, then one
would simply work with the algebra
\[ \CA= p M_N(C^\infty_0(\CQ) )p\]
where $p\in M_N(C^\infty_0(\CQ) )$,  $p^2=p$  and tr$p\,=n$.
\end{remark}

\subsection{The two point model}
We now come to another classic in NCG, the two-point-space, i.e.
\[ \CA = C^\infty_0(\CQ)\otimes(\C\oplus\C) .\]
Algebra elements $a\in\CA$ can then be viewed as diagonal
two-by-two matrices
\[ a = \left(\begin{array}{cc} a_1 & 0 \\ 0 &a_2 \end{array} \right) \]
with entries $a_1,a_2$from $C^\infty_0(\CQ)$ . We shall represent
$\CA$ on the space $\CH=L^2(\CQ,E) \otimes \C^2$. In this case,
the commutant of the algebra equals the (weak closure of the)
algebra itself
\[  \CA' = \CA  .\]
Since $\CA = \CZ = \CA^o$ is commutative, our NC-Uncertainty
agrees with the commutative one. Once again, it will tell us that
\[ \dot{b}_t = \left(\begin{array}{cc} X_{b_1} & 0 \\ 0 &X_{b_2}
\end{array} \right) +A(b) \]
where $A(b)$ lies in the commutant of $\CA$, i.e. in $\CA$ itself.
Moreover, the map $b\to A(b)$ is linear and fulfills the Leibniz
rule. Thus there will exist one forms $A_i,\alpha_i$ such that
\[ A(b) =\left(\begin{array}{cc} g_1(A_1,{\rm d}b_1) + g_1(\alpha_1,{\rm
d}b_2) & 0 \\ 0 & g_2(A_2,{\rm d}b_2) + g_2(\alpha_2,{\rm d}b_1)
\end{array}
\right)
\]
where $g_1,g_2$ denote the metrics on the different world sheets.
The terms $g_1(\alpha_1,{\rm d}b_2)$ (and resp. ...) seem to carry
particles from one world sheet to the other. However, there are not
present, i.e., $\alpha_k=0 $! In fact, if one decomposes the
Hamiltonian according to
\[ H= \left(\begin{array}{cc} h_{11} & h_{12}\\ h_{21} & h_{22} \end{array}
\right) .\] Then one easily computes
\[ [H,a] = \left(\begin{array}{cc} [h_{11},a_1] & h_{12}a_2 -a_1h_{12} \\
h_{21}a_1 - a_2h_{21} & [h_{22},a_2] .
\end{array}
\right) \] Thus, a term like the one above can never be generated
by a Hamiltonian.
\\

\noindent Nevertheless  one might like to construct a model which
"mixes the two manifolds". In order to see what has been wrong
with our assumptions, let us consider such a toy Hamiltonian
\[  H = \left( \Delta(g,A)  + V \right) \ei_2   + \left(\begin{array}{cc} 0
& \phi \\
\bar{\phi} & 0 \end{array} \right) \]  where $\phi$ is a complex
scalar field. Such a term could be reminiscent of  spontaneous
breaking via a Higgs field of $U(1)\times U(1)$ (which of course,
does not happen here -- it can only happen in a quantum field
theory...we consider our theory to arise as a limit....). Then,
after a short computation one gets
\[  [a,\dot{b}] = ig({\rm d}a , {\rm d}b) \ei_2+ i\left(\begin{array}{cc} 0
& \phi (a_1-a_2)(b_1-b_2)
\\ \bar{\phi}(a_2-a_1)(b_2-b_1) & 0 \end{array} \right) .\]
So, such a term would violate the canonical uncertainty relation,
which might therefore be a too strong requirement.

\subsection{The noncommutative double-torus}

Let us now come to a much more noncommutative, yet still very
simple example. The algebra we shall consider can be viewed as the
bi-cross product of the two-torus $C(T^2)$ with the diffeomorphism
that is not generated by a vector field, i.e.
the one which interchanges the two circles in $T^2=S^1\times S^1$.\\
Thus $\CA$ is generated by two mutually commuting unitaries $U,V$
and one additional generator $\sigma$, subject to the relation
\[ \sigma U = V\sigma , \qquad\qquad \sigma^* = \sigma \qquad\qquad \sigma^2=1 .\]
Note that every algebra element $a\in\CA$ can be written in the
form $a=f_- + f_+ \sigma $, where $f_\pm$ are power series in $U$
and $V$, i.e. they van be identified with functions over the torus
$T^2$. The product in the algebra $\CA$ is then given as
\[ (f_- + f_+ \sigma)(g_- + g_+ \sigma) = (f_-\cdot g_- + f_+\cdot {\tilde g}_+) +
(f_-\cdot g_+ + f_+\cdot {\tilde g}_-)\sigma .\] where the dot
``$\cdot$'' denotes the usual point-by-point product of functions
over $T^2$ and ${\tilde g}_\pm = \sigma g_\pm \sigma$. As is
easily seen the center $\CZ$ of the algebra $\CA$ is given by
elements of the form $f + \sigma f\sigma $
with $f\in C(T^2)$. Thus, not expected, the center consists of functions on the torus which are invariant under $\sigma$.\\

This algebra can then be represented on the Hilbert space
$L^2(T^2)\oplus L^2(T^2)$, with the standard orthonormal basis
\[ |n,m,\pm \rangle \qquad\qquad \qquad n,m \in \mathbb{Z} \]
as follows
\begin{eqnarray*}
U |n,m,\pm \rangle & = & |n+1,m,\pm \rangle \\
V |n,m,\pm \rangle & = & |n,m+1,\pm \rangle \\
\sigma |n,m,\pm \rangle & = & |m,n,\mp \rangle .
\end{eqnarray*}
Note that the vector $|0,0, -\rangle$ cyclic and separating. We
can thus apply the Tomita-Takesaki Theorem to obtain
\begin{eqnarray*}
J |n,m,+ \rangle & = & |-n,-m,+ \rangle \\
J |n,m,- \rangle & = & |-m,-n,- \rangle ,
\end{eqnarray*}
and the corresponding ``opposite algebra'':
\begin{eqnarray*}
U^o |n,m,+ \rangle & = & |n+1,m, + \rangle \\
U^o |n,m,- \rangle & = & |n,m+1,- \rangle \\
V^o |n,m,+ \rangle & = & |n,m+1,+ \rangle \\
V^o |n,m,- \rangle & = & |n+1,m, - \rangle \\
\sigma^o |n,m,\pm \rangle & = & |n,m,\mp \rangle .
\end{eqnarray*}
We can now start to investigate the consequences of the strong
NC-Uncertainty relation $[\dot{a}_t, b_t^o] \in \CA\otimes\CA^o$ for
families of representations of $\CA$ which are all unitarily
equivalent to the one described above, the unitaries being generated
by a Hamiltonian $H$. For simplicity we shall assume that $H$ is
time independent and commutes with J,
\[ [H,J]=0 \qquad \Rightarrow \qquad H|n,m,\pm \rangle =  h_{\pm} |n,m,\pm \rangle \]
with two operators $h_\pm$ obeying $h_+=\bar{h}_+$ respectively $\bar{h}_- = \sigma h_- \sigma$.\\

A lengthy but straightforward calculation of the  commutators
$[\dot{U},U^o]$, $[\dot{U},V^o]$ $[\dot{V},V^o]$ then shows that
they are in $\CA\otimes\CA^o$ if and only if $h_+ = h_- =: h$ and
such that $[h,a]$ is a derivation on $\CA$ for all $a$ that are
power series in $U$ and $V$, i.e. \[ [[h,a], b] \in
\CA\qquad\qquad\qquad \forall b\in \CA .\] Note that it then
immediately follows that $h=\sigma h \sigma$ and hence that \[
\dot{\sigma} = [H,\sigma] = 0. \] In order to obtain the most
general Hamilton-Operator it is convenient to introduce the two
standard vector fields on the torus which generate the $U(1)\times
U(1)$ -Symmetry,
\begin{eqnarray}
\delta_1  |n,m,\pm \rangle & =& n |n,m,\pm \rangle \\
\delta_2  |n,m,\pm \rangle & =& m |n,m,\pm.\rangle
\end{eqnarray}
Note that they are not derivations on the algebra, since they are
not compatible with the relation $U\sigma = \sigma V$. However,
the combination $P = z (\delta_1 + \delta_2)$ is a derivation on
$\CA$ whenever
$z\in \CA$, i.e. whenever $z$ is a function on the torus that is invariant under $\sigma$.\\
One then easily infers that the most general Hamilton-Operator is
of the form
\[ H = z_1(\delta_1^2 + \delta_2^2) + z_2 \delta_1\delta_2 + z_3 (\delta_1+\delta_2) + z_4 \sigma \qquad\qquad
z_i\in \CZ\quad i\in \{1,2,3,4\} .\] The interesting piece here is
the term $z_4\sigma$ which, since the diffeomorphism $\sigma$ is
not generated
by a vector field is highly nonlocal on $T^2$.\\

However, the other terms apart from $z_4\sigma$, have a rather
clear interpretation: \\
The above construction is, in fact a standard technique in
noncommutative geometry. Given the algebra $C^(T^2)$ and the
diffeomorphism $\sigma$ one may construct the algebra $C^(T^2)/\sim$
of equivalence classes under the equivalence relation given by
$\sigma$ (i.e. $f \sim g \Leftrightarrow f \sigma g \sigma$.)
Obviously $C^(T^2)/\sim$ is isomorphic to $\CZ$ from above. But it
is also {\em Morita-equivalent} to the full algebra $\CA$. Thus, in
order to do some differential topology on $C^(T^2)/\sim$ one may
also use $\CA$ to obtain the
same invariants.\\
But the above Hamilton Operator is just the most general Laplace
Operator on the spectrum of $C^(T^2)/\sim = \CZ$ plus the highly
nonlocal term $z_4\sigma$ whose presence is due to the fact that
the algebras are "only" Morita equivalent but not isomorphic.

\section{Noncommutative Tori}\label{nc-tor}
Finally we examine the noncommutative torus, i.e., the algebra
generated by two unitaries $U_1, U_2$ subject to the relation
\[ U_1 U_2 = \lambda U_2 U_1, \qquad\qquad \qquad \lambda=e^{i2\pi\theta} . \]
More precisely algebra elements $a$ are power series $a \sum_{kl} a_{kl} U_1^kU_2^l $ with coefficients $a_{kl}$ which
vanish faster than any polynomial for $k,l \to\infty$.
\\
Without loss of generality, we assume $\theta$ being irrational.
It is well known that for $\theta$ rational the algebra $\CA$
isomorphic to the algebra of endomorphisms of the space of
sections of a vector bundle over the commutative torus, i.e. it is
of the form $p M_n(c^\infty_0(T^2)) p$. Thus for rational $\theta$
we would be back to the almost commutative case already described.
\\
We would like to treat the noncommutative Torus as a deformation
of the commutative one which
corresponds to $\theta=0$ .\\

Now, it is well-known \cite{ElementsNCG}, \cite{BEJ}, that
derivations $P$ on the noncommutative Torus are of the form
\[ P = \sum_i c_i \delta_i + a \qquad \qquad\qquad a\in \CA, \qquad c_i\in
\C \] where $\delta_i$ are the standard derivations
$$\delta_i U_j = \delta_{ij} U_j . $$

\noindent We have not yet said anything about the representation
of $\CA$. We take
 $\CH= L^2(T^2)$, with its basis $|n_1,n_2\rangle $, $n_k\in \Z$. The
representation (that is also used for the construction of he
spectral triple...) is then given by \beas
U_1 |n_1,n_2 \rangle & = & \lambda^{n_2}  |n_1+ 1,n_2 \rangle \\
U_2  |n_1,n_2 \rangle & = &  |n_1,n_2 +1\rangle . \eeas Note that
this representation possesses a cyclic, separating vector, namely
$ |0,0 \rangle$. According to the Tomita-Takesaki-Theorem there
then exists an antiisomorphism $J$ from $\CA$ to its commutant
$\CA'$, which is therefore isomorphic to "opposite algebra"
$\CA^o$, i.e. $\CA$ equipped with the reversed product $ a^o b^o (ba)^o $. Explicitly one finds that $\CA' = \CA^o$ is generated by
\beas
U_1^o |n_1,n_2 \rangle & = &   |n_1+ 1,n_2 \rangle \\
U_2^o  |n_1,n_2 \rangle & = & \lambda^{n_1} |n_1,n_2 +1\rangle .
\eeas In the commutative case ($\lambda=1$) obviously $\CA=\CA'$.
However, for irrational $\theta$ the situation is completely
different. Then the center of $\CA$ is trivial, and thus \[\CA\cap\CA´=\C\ei.\] It is therefore natural to replace the algebra
$\CA$ by the larger algebra $\CA\otimes\CA'$. The resulting
algebra has trivial center. We therefore need a reasonable
candidate for the uncertainty relation. Before we come to that, it
is helpful to make the following observation:\\ The standard
derivations $\delta_k$ are represented on $\CH$ as
\[ -i\delta_k |n_1,n_2 \rangle = n_k |n_1,n_2 \rangle. \]
One then calculates:
\[  [-i\delta_k, U_l^o] = \delta_{kl} U_l^o  \]
The algebra $\CA^o$ does possess, of course, the same derivations
as $\CA$, i.e. only inner derivatives and complex linear
combinations of the standard derivations $\delta_k$.
\\

\begin{proposition}
Assume that $\frac 1 {\mid 1-\lambda^n\mid} = O(n^k)$ for some $k$
(Note, that this is satisfied by generic $\lambda$, while those
$\lambda$ not satisfying this condition are of measure zero). Then
any derivation $\delta: \CA\rightarrow \CA\otimes\CA^o$ from $\CA$
into the $\CA$-bimodule $\CA\otimes\CA^o$ can be uniquely decomposed
into
\begin{align}
\delta := c^i\delta_i+\tilde{\delta},
\end{align}
where $c^i$ are in $\CA^o$, $\delta_i$ are the standard
derivations on $\CA\cong \CA\otimes 1$ and $\tilde{\delta}$ is an
inner derivation on $\CA\otimes\CA^o$.
\end{proposition}

\proof We extend and follow closely the proof given in \cite{BEJ}.

Any derivation $\delta$ can be given by its action on the
generators $U_i$:
\begin{align}\label{deriv}
\delta (U_1) &= \sum_{n,m,p,q}{c^1_{mnpq} U_1^{n+1}U_2^m {U_1^o}^{p}{U_2^o}^{q}}\\
\delta (U_2) &= \sum_{n,m,p,q}{c^2_{mnpq} U_1^{n}U_2^{m+1}
{U_1^o}^{p}{U_2^o}^{q}}
\end{align}
with the coefficients $c^i_{mnpq}$ being in Schwartz space and
satisfying consistency relations due to the commutation relations
between $U_1$ and $U_2$:
\begin{align}
c^1_{mnpq}(\lambda^{-n}-1) + c^2_{mnpq}(\lambda^{-m}-1) =0
\end{align}
An inner derivation of $\CA\otimes \CA^o$ is formally given by the
commutator with $$ B = \sum_{m,n,p,q}{d_{mnpq}U_1^{n}U_2^m
{U_1^o}^{p}{U_2^o}^{q}},$$ where the coefficients $d_{mnpq}$ have to
satisfy the following to match (\ref{deriv}):
\begin{align}
d_{mnpq}&\begin{cases}
\frac {c^1_{mnpq}} {\lambda^{-m}-1} \text{ if $m\neq 0$} \\
\frac {c^2_{mnpq}} {1-\lambda^{-n}} \text{ if $n\neq 0$}
\end{cases}
\end{align}

Thus, an inner derivation can provide for all coefficients except
for $c^i_{00pq}$ which has in this case to be set to be zero,
$$c^i_{00pq}=0.$$
However, these are exactly given by linear combinations of the
standard derivations with coefficients ${U_1^o}^{p}{U_2^o}^{q}\in
\CA^o$. Thus, any derivation can be uniquely decomposed into a
combination of standard ones with coefficients in $\CA^o$ and into
an inner part.

The convergences of series in the formal argument above can be
checked following \cite{BEJ}.

From the strong NC-uncertainty condition we know that the
derivation $[\bullet, [b^o,H]]:\CA\rightarrow \CA\otimes\CA^o$ can
be decomposed as above:

\begin{align}
[\bullet, [b^o,H]]&= d^i(b^o)\delta_i + [\bullet, A_0],&\text{
with $A_0\in \CA\otimes\CA^o$, $d^i(b^o)\in\CA^o$.}
\end{align}
It is easily checked that $d^i: \CA^o\rightarrow \CA^o$ is a
derivation and can be decomposed accordingly:
\begin{align}
d^i &= c^{ij}\delta_j + [\bullet, A^i] &\text{ with $A^i\in
\CA\oplus\CA^o$, $c^{ij}\in\mathbb{C}$.}
\end{align}

Thus, the general Hamiltonian is given by

\[ H = \oh \sum_{kl} c_{kl} (\delta_k - A_k)(\delta_l-A_l) +A_0, \]
where $c_{kl}\in \C$, $A_k\in \CA\oplus\CA^o$ and $A_0\in \CA
\otimes\CA^o$. We would first like to point out, that the above
Hamiltonian coincides (if $A_0=0$) with the most general Laplace
operator, that one get from Connes' notion of spectral triples: If
one takes the square of the most general Dirac-Operator on the
noncommutative torus \cite{ElementsNCG} one obtains precisely the
above. Then, however one would have have to assume that $H$
commutes with $J$.\\
More explicitly, the Dirac-Operator, acting on the space
$L^(T^2)\otimes\C^2$ is given as
\[ D= \sum\limits_{i=1}^2\sigma^i(\delta_i -a_i-a_i^{o})\]
where $\sigma^1$ denote selfadjoint two by two matrices such that
$\sigma^i\sigma^j +\sigma^j\sigma^i = 2c_{ij}$  and the $a_i$ are
selfadjoint algebra elements. One then has obviously $D^2=H$
with $A_k=a_k+a_k^o$, as required by commutation with $J$.\\
At first sight, it may appear as if this Dirac-operator describes
the flat metric $c_{ij}$ on the noncommutative torus. Yet, in view
of Connes distance formula for two states $\chi,\varphi$ on $\CA$:
\[ d(\chi,\varphi) = \sup\limits_{\|[D,a]\|\leq1}\,\,\left\{ |\chi(a)-\varphi(a)| \right\}  \]
this is not necessarily true: As the algebra elements $a_i$ do not
commute with algebra elements, they do affect the set over
which the supremum is to be taken, and thus the distance itself.
Thus the metric is affected by the choice of the gauge potential
$A_k$. This, however, is not unexpected. The gauge potential
refers to inner automorphisms of the algebra, while the metric (or
rather the Levi-Civita connection) gauges the outer automorphisms.
But the noncommuative torus essentially has only inner
automorphisms, and hence one would expect that curvature only
arises from the gauge potential. Note however, that the freedom in
the choice of $A_k$ is very large -- unlike the commutative case,
where it would in fact be zero if it had to commute with $J$.

\section{Conclusions and Outlook}
In this  paper, we have presented an axiomatic framework for
nonrelativistic quantum mechanics on noncommutative configuration
spaces. One of the  main virtues of our approach as compared to
approaches which work with the ``canonical momentum'' $p_k$ which is
usually taken as translation operators, and thus in particular as
derivations on the algebra $\CA$, is that it is not restricted to
algebras which possess such derivations. Moreover, the velocities
$\dot{x}_k$, that we laid emphasis on,  do have a clear physical
meaning and one may immediately propose experiments to measure them,
in contrast to the momentum operators  $p_k$ which are usually
measured via the velocities.

More importantly, our approach does not depend on the choice of coordinates. As we have pointed out, this is of particular importance on
noncommutative spaces.\\

As we have shown in several examples, our generalized uncertainty
relations do, despite their rather general appearance, pose tight
restrictions on the dynamics
 of such a theory. In particular, on the Moyal deformed plane we find only such Hamiltonians which would correspond
to a flat metric. (The same would hold in three dimensions). Our
uncertainty relation is the equivalent to the usual description via
the canonical momenta $[p_k,x_l]= i\hbar \delta_{kl}$. They are , however, valid in any coordinate system. \\
We should stress that modified uncertainty relations of the form
$$ [x_k,\dot{x}_l]=i\hbar \delta_{kl} + O(\frac{1}{p^2})$$ which  are found for instance in string theory
can also be discussed in our framework, as the modification of the
usual uncertainty relation is given by a compact operator. However,
we have not found a convincing route to the most general Hamiltonian
for such a relation. They are thus left for future work. More
importantly however, we shall then try to set up scattering theory
in our framework, as that might lead to definite experimental
predictions.

\section*{Acknowledgements}
The authors would like to thank for support from the Alexander von
Humboldt Foundation and from grants MSM: J10/98:192400002,
MSM4781305904 and GA\v{C}R 202/05/2767.

\end{document}